\documentclass[aps,prl,twocolumn,showpacs,superscriptaddress,groupedaddress]{revtex4-1}
\usepackage{graphicx}
\usepackage{dcolumn}
\usepackage{bm}
\usepackage{amssymb}
\usepackage{amsmath}
\usepackage{subfigure}

\begin{document}\par
\title{Quantum Dynamics of a Nanomagnet driven by Spin-Polarized Current}
\author{Yong Wang and L.J. Sham}\email{lsham@ucsd.edu}
\affiliation{Center for Advanced Nanoscience, Department of Physics,
University of California, San Diego, La Jolla, California 92093-0319, USA}
\begin{abstract}
A quantum theory of magnetization dynamics of a nanomagnet as a sequence
of scatterings of each electron spin with the macrospin state of the
magnetization results in each encounter a probability distribution of
the magnetization recoil state associated with each outgoing state of
the electron. The quantum trajectory of the magnetization contains the
average motion tending in the large spin limit to the semi-classical
results of spin transfer torque and the fluctuations giving rise to
a quantum magnetization noise and an additional noise traceable to the
current noise.
\end{abstract}
\maketitle
\emph{Introduction} The spin transfer torque (STT) \cite{STT1,STT2} is both a fundamental
process in magnetization dynamics and important to spintronics in memory and
information processing integration. The magnetization dynamics driven by
spin-polarized current via STT has made great progress by experiments and by
a ``semiclassical'' theory of treating the magnetization dynamics
classically and the polarized current transport and the spin waves quantum
mechanically (see a series of reviews introduced by Ref.~\cite{mdym}).

Recent development in fast time-resolved measurements \cite{ohno08pulse,tomita08,cui10}
and coherent control\cite{webb} have made possible studies of magnetization dynamics
and fluctuations not masked by the inhomogeneity effects of the measurement and the prospect of precision
control. These recent experiments including Ref.~\cite{krivo10}, by showing
the stochastic nature of the magnetization dynamics at short time intervals,
highlight three important aspects of the subject. First is the stochastic
motion which may lead to an understanding of noise, a fundamental issue in
magnetization dynamics \cite{suhl}. Second, in interconnected systems at the
nanoscale, the particulate nature of the current electrons imprints shot
noise on the magnetization through the quantum scattering process \cite{foros,chud}.
 Third, the thermal fluctuations are important because, for
example, the thermal noise power is comparable to the microwave power
generated by the magnetization precession \cite{slavin}. The measurements
seem to be satisfactorily treated by the semiclassical theory including
micromagnetics. However, the common treatment of angular momentum transfer
by the magnet as an immovable scattering potential is shown below by order
of magnitude estimate to be inadequate for accurate computation of mean
displacement and fluctuations in a nanomagnet. The origin of the stochastic
motion arises from the assumption of randomness in either the current or the
magnetization. The spin waves seem grafted on rather arising naturally out
of the vibration of the spins which constitute the moving magnetization.
These points motivate us to question if inclusion of the quantum dynamics of
the magnetization may not only provide more accurate connection between the
current electrons and the magnetization but also provide qualitative
understanding of the stochastic dynamics as well as bringing out the
prospect of coherent motion for precision control. By a quantum theory of
the nanomagnet which gives its mean dynamics and fluctuations through
scattering between the current electrons and the movable magnet, we hope to
illustrate the fundamental aspect of our approach to STT. The development of
quantum optics after the laser operation was understood by semiclassical
theory provides perhaps an optimistic historical guide for the development
of coherent magnetization dynamics after the successes of the semiclassical
STT theory.

\emph{Conditional Scattering and Stochastic Schr\"{o}dinger Equation} We take
the magnetization of a single-domain nanomagnet to be represented by
a coherent state of the local spins \cite{Cohe} and view its dynamics in the
spin-polarized current as a sequence of scattering by individual electrons.
Since each scattering produces a number of outgoing macro-spin states, the serial
scatterings create a Monte Carlo trajectory of the macro-spin states which is
governed by a stochastic Sch\"odinger equation \cite{carmichael}. Repeated solutions form an ensemble
of quantum trajectories for the magnetization from which the mean trajectory
and the fluctuations can be obtained. Since the outcomes of the macro-spin
states are paired with the scattered spin states of the current electron to
form an entangled state, the scattered electron state may be viewed as
conditioned on the corresponding macro-spin state.
After the first scattering, the electron recedes and the magnetization is
left in a reduced density matrix of the outgoing macro-spin states. The second
electron from the current may be viewed as the recipient of one of the macro-spin states.
This step is equivalent to measurement or decoherence\cite{wiseman},
and the magnetization undergoes a Brownian motion and not a quantum random walk.
The magnetization noise then has three sources, an intrinsic one due
to the uncertainty of the quantum state of the magnetization, two extrinsic
ones due to the entangled state after each scattering and due to the
particulate nature of each colliding electron with the coherent macro-spin
state. The last forms a channel for the mutual effects of the current noise
and the magnetization noise.

We illustrate the quantum effects with a simple scattering model of the
nanomagnet by a regular sequence of electrons and then layer the
complexities of the real current\cite{supp}. We study here the dynamics of the
magnetization as an angular momentum coherent state $|J,\Theta,\Phi\rangle$
\cite{Cohe}. The effect of the spin waves as small deviations from the
coherent state is being investigated. $J$ is the quantum number of the total
spin $\mathbf{\hat{J}}$ of the localized $d$-electrons in the nanomagnet and
is of the order of $10^{6}$ for typical nanomagnets. The angles $\Theta$ and
$\Phi$ characterize the direction of the total spin of the nanomagnet. The
current electron has the incoming state $|k,\uparrow\rangle$, its wave
vector $k$ impinging normally on a ferromagnet film in the $x=0$ plane and
its spin along the $z$-axis. The scattering potential, $\delta(x)[\lambda_{0}
\mathsf{\hat{J}}_{0}+\lambda\mathbf{\hat{s}}\cdot\mathbf{\hat{J}}]$, is a
potential of fixed position but includes a spin independent term indicated
by the identity operator $\mathsf{\hat{J}}_{0}$ of the total spin state
space and the exchange term between the current electron spin and the total
local spin. The $\delta$-potential models the FM film of thickness of
several nanometer, and the parameters $\lambda_{0}$ and $\lambda$ are the
interaction strength between the current electron and one $d$-electron,
estimated from two simple exchange-split potentials.

\begin{figure}[tbp]
\subfigure{
    \begin{minipage}{0.1\textwidth}
    \centering
 \includegraphics[scale=0.23]{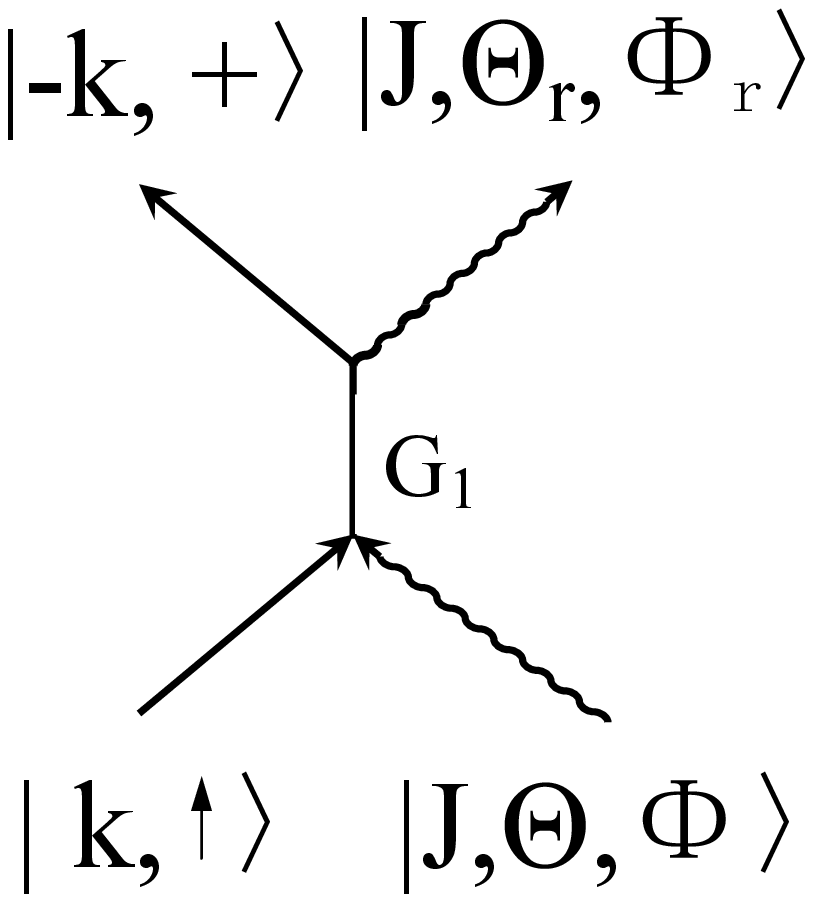}
    \end{minipage}}
\subfigure{
    \begin{minipage}{0.1\textwidth}
    \centering
  \includegraphics[scale=0.23]{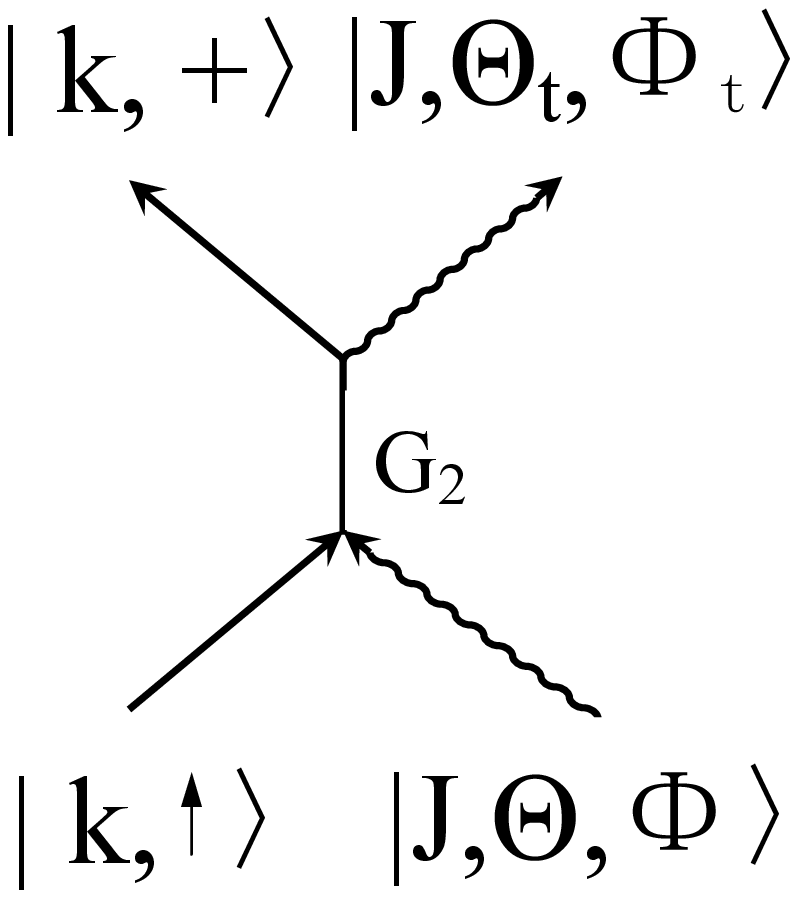}
   \end{minipage}}
\subfigure{
    \begin{minipage}{0.1\textwidth}
    \centering
 \includegraphics[scale=0.23]{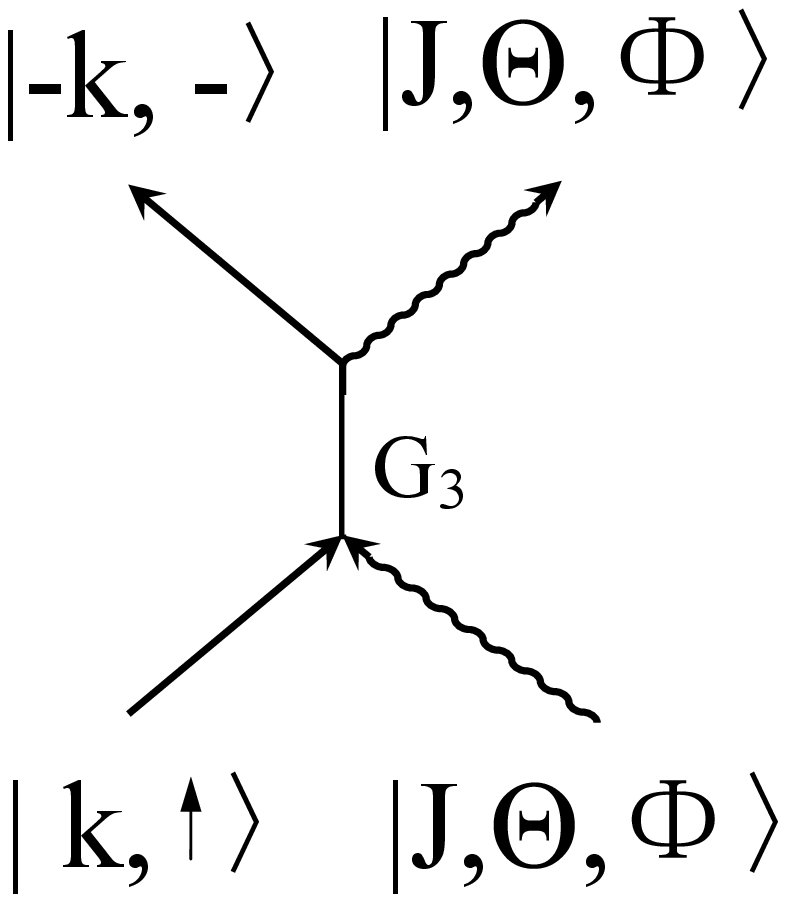}
    \end{minipage}}
\subfigure{
    \begin{minipage}{0.1\textwidth}
    \centering
  \includegraphics[scale=0.23]{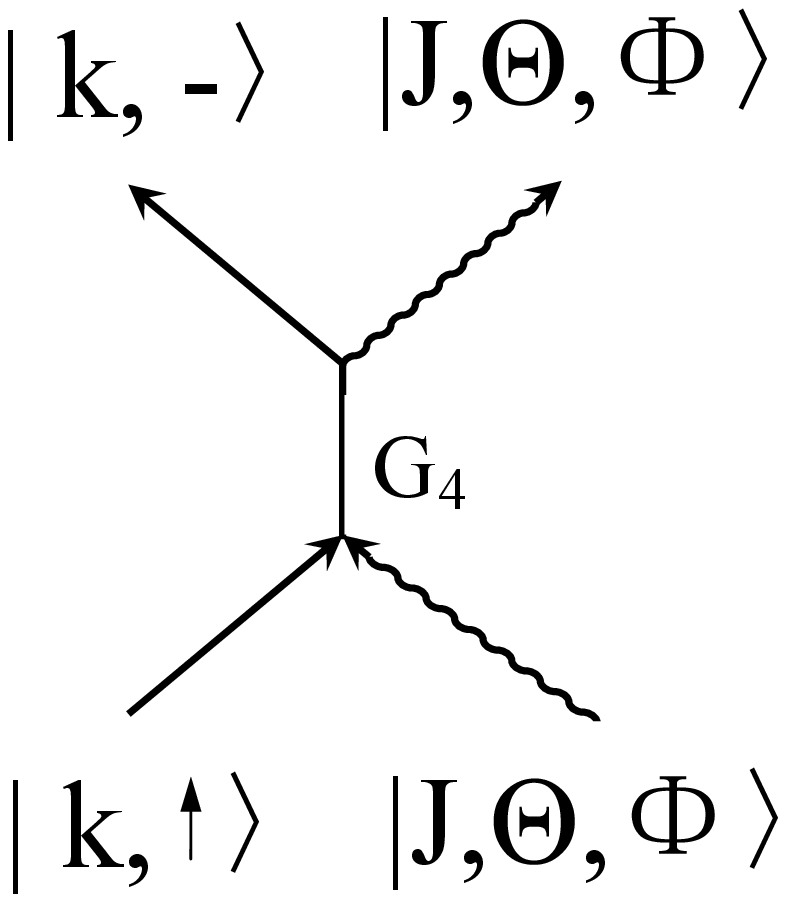}
  \end{minipage}}
\caption{Four possible states of the nanomagnet in the Berger basis $
\{|\pm k,\pm\rangle\}$ after scattering. The straight lines represent the
spin-polarized electron and the wavy lines represent the nanomagnet states.
The macro-spin state $|J,\Theta_i,\Phi_i\rangle$ with suffices $i =$1---4
define the association with the electron spin states in the Berger basis.
$G_{i}$ are the conditional probabilities.}
\label{FeynDiagram}
\end{figure}

After a scattering event by one current electron, the total state of the
system of the nanomagnet and the current electron is entangled\cite{supp}. In the
Berger \cite{STT2} basis states $\{|\pm k,\pm\rangle\}$ for the current
electron, in which the $z$ axis is rotated in the $z$-$\mathbf{J}$ plane to
the magnetization direction before the scattering, the conditional
macro-spin states associated with the outgoing single electron states are
coherent states with an error of the order $1/J^2$, \vspace*{-0.07in}
\begin{equation}
|\Psi\rangle= \sum_{i=1}^4 g_{i}
|J,\Theta_{i},\Phi_{i}\rangle|k_i,\sigma_i\rangle,  \label{entanwav}
\end{equation}
where Span$(|k_i, \sigma_i\rangle) = (|-k, +\rangle, |k, +\rangle,
|-k,-\rangle, |k,-\rangle)$. The Berger basis is therefore preferable to the
original spin basis of the current electron of spin up and down along the $z$
-axis, and only the incoming electron state is shown in the original basis.
The Berger basis may be viewed as constituting a measurement basis for the
current electron with the outcomes associated with the probabilities $
G_{i}=|g_{i}|^{2},i =$ 1---4. A conditional state of the macro-spin state of
the magnetization is determined by the scattered current electron state, as
shown by the Feynman diagrams in Fig.~\ref{FeynDiagram}. Note that, after
scattering, the conditional macro-spin states associated with the electron
states $|\pm k,-\rangle$ suffer only a phase change because the flip-flop
terms $\widehat{s}_{+}\widehat{\mathsf{J}}_{-}+\widehat{s}_{-}\widehat{
\mathsf{J}}_{+}$ in the Heisenberg exchange interaction do not connect the
state $|k,+\rangle|J,J\rangle$ to $|k,-\rangle|J,m\rangle$ for any $m<J$.

The evolution of the quantum state of magnetization from $|J,\Theta,\Phi\rangle$
to $|J,\Theta_i,\Phi_i\rangle$ stochastically by one spin-polarized current electron
may be represented by a rotation, $|J,\Theta^{\prime},\Phi^{\prime}\rangle=e^{-i\mathbf{\hat{n}}\cdot\mathbf{\hat{J}}
\vartheta}|J,\Theta,\Phi\rangle$, with the rotation axis $\mathbf{\hat{n}}
=(-\sin\varphi,\cos\varphi,0)$ and the rotation angle $\vartheta$ as the
random variables with probability outcomes determined by Eq.~(\ref{entanwav}).
 Since the angle $\vartheta$ is of the order of $1/J$ and the time
interval $\tau$ between two successive electrons is small compared with the
time scale of magnetization dynamics, the stochastic rotation of the
nanomagnet may be treated as a continuous rotation governed by the
stochastic Schr\"{o}dinger equation,
\begin{eqnarray}
i\frac{\partial}{\partial t}|J,\Theta,\Phi\rangle=\omega\mathbf{\hat{n}}\cdot
\mathbf{\hat{J}}|J,\Theta,\Phi\rangle,  \label{SSch}
\end{eqnarray}
under the effective magnetic field due to the polarized current electrons,
producing precession frequency, $\omega\equiv\vartheta/\tau$ about the axis $
\mathbf{\hat{n}}$.

\begin{figure}[tbp]\flushleft
\includegraphics[scale=0.32]{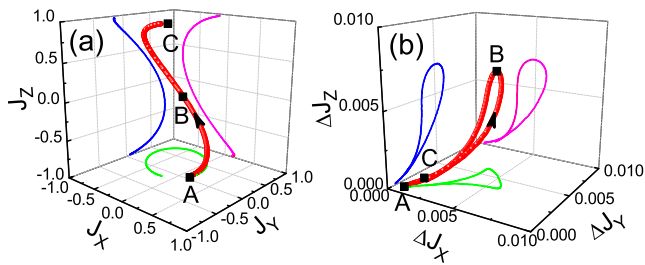}
\caption{(color online). Time evolution of magnetization and its noise. (a) The
average trajectory of the macro-spin state under STT and its projections
onto the Cartesian planes. The Cartesian components $\mathsf{J}_{\protect
\alpha}, \protect\alpha = x,y,z$ are normalized in units of $J$. (b) The
time evolution of the uncertainties of the macro-spin components and its
projections onto the Cartesian planes. $\Delta\mathsf{J}_{\protect\alpha}$
are normalized in units of $J$. The current electron number is $
N_{e}=1.5\times 10^{7}$ and the macro-spin $J=10^6$. The initial orientation
is $(\Theta_{0}, \Phi_{0}) = (3, 0.5)~$rad. The electron wave vector obey
the probability function $f(k)=2k/k_{F}^{2}$ for $k\in[0,k_{F}]$ with $
k_{F}=13.6$~nm$^{-1}$, and the spin polarization vector is taken $\mathbf{S}
=(0,0,1)$. $\protect\lambda_{0}$ and $\protect\lambda$ are determined by the
spin-dependent potential $\Delta_{+}=1.3$~V, $\Delta_{-}=0.1$~V, and layer
thickness $d=3$~nm. The switching time between points $A$ and $C$ is
estimated as $1.3$~ns for bias $V=1$~mV and cross-section area $\mathcal{A}=10^{4}$~nm$
^{2}$, by Eq.~(\protect\ref{currandnoise}).}
\label{Mag}
\end{figure}

\emph{Dynamics and Fluctuation of Magnetization, Electric Current and Current Noise}
To study the switching behavior of nanomagnet at zero magnetic field, we
computed an ensemble of 500 runs of Monte Carlo steps, each with sequential
scatterings by $N_{e}=1.5\times 10^{7}$ electrons with the macro-spin $J=10^6 $.
To allow for the three dimension nature of the electron wavevector
distribution in a Fermi sphere of radius $k_F$ in the normal metal model,
the normal component to the interface $k$ obeys the distribution function $
f(k)= 2k/k_{F}^{2}$ for $k\in[0,k_{F}]$\cite{supp}. In our simulation, the wavevector $
k $ of each injected electron is generated randomly according to $f(k)$. The
scattered state in each step is randomly selected according to the
probabilities $G_{i}$, associated with the macro-spin state $
|J,\Theta_i,\Phi_i\rangle$ and output electron state $|k_{i},\sigma_{i}
\rangle$. The effect of STT comes from the random encounter by each incoming
electron of a choice of the macro-spin states left by the previous electron,
giving rise of a random walk of the magnetization. Fig.~\ref{Mag} shows the
direction of the magnetization $\mathbf{\mathsf{J}}$ averaged over the
ensemble of trajectories and its uncertainties in the Cartesian components $
\Delta \mathsf{J}_\alpha$. 500 runs were found to be sufficient to stabilize
the ratio of the uncertainties to the average.

Fig.~\ref{Mag}(a) shows that the switching motion also contains a precession
about the spin polarization direction of the injected electrons, which
reproduced the semiclassical results of being driven by
$\mathbf{M}\times\mathbf{S}$ and $\mathbf{M} \times(\mathbf{M}\times\mathbf{S})$
for the magnetization $\mathbf{M}$ and current spin polarization $\mathbf{S}$.
It is also found that the number of
electrons needed for switching is of the order of $J$. For typical
experiments \cite{nanomag}, the electric current is of the order of mA and
the switching time is of the order of ns, both consistent with the electron
number $N_e$ we used. Since the input parameters to our model, namely the
Heisenberg exchange energy and the current, are of the order of the physical
properties of real systems, we see the ball-park agreement of the switching
time as an encouraging sign for the quantum approach so that the noise
effects from the theory may be worth being included with the thermal noise
in interpretation of experimental measurements. Without the thermal noise,
the fluctuations $\Delta\mathsf{J}_{\alpha}$, shown in Fig.~\ref{Mag}(b) in
time order from point $A$ to $C$, first increase till the midpoint $B$ and
then decrease to $C$. The initial state of nanomagnet is a pure angular
momentum coherent state $|J,\Theta,\Phi\rangle$ with only intrinsic
magnetization noise ($\Delta\mathsf{J}_{\alpha}\sim\sqrt{J}$). Scattering by
current electrons leads to stochastic motion which accounts for the rise of
the extrinsic magnetization noise. As all the trajectories converge to the
state $|J,0,0\rangle$, the magnetization noise finally decreases to the
intrinsic quantum fluctuation.

\begin{figure}[tbp]
\flushleft
\includegraphics[scale=0.32]{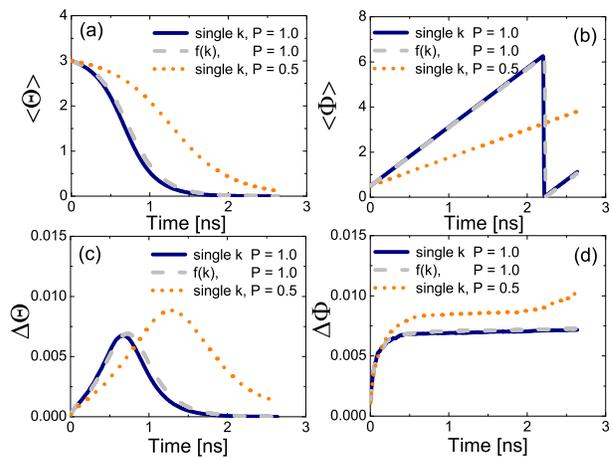}
\caption{(color online). Effects of injected current electron noise on the
magnetization dynamics of nanomagnet. (a) Time evolution of the average
value of $\Theta$. (b) Time evolution of the average value of $\Phi$. (c)
Time evolution of the fluctuation of $\Theta$. (d) Time evolution of the
fluctuation of $\Phi$. Three cases are considered here: the fully polarized
electrons with a single wave vector $2k_{F}/3$ (solid line); the fully
polarized electrons obey wave vector distribution function $
f(k)=2k/k_{F}^{2} $ for $k\in[0,k_{F}]$ (dash line); partially polarized
electrons with a single wave vector $2k_{F}/3$ (dotted line). The spin
polarized vector is $\mathbf{S}=(0,0,P)$. The current electron number is $
N_{e}=3.0\times10^{7}$, and the other simulation parameters are the same as
in Fig.~\protect\ref{Mag}. }
\label{noiscomp}
\end{figure}

The injected current electron noise can come from either its wave vector
distribution $f(k)$ or the uncertainty of its spin state. To identify their
effects on the magnetization dynamics of nanomagnet, we include them
separately in the simulations and compared the case with no injected electric
current noise. As shown in Fig.~\ref{noiscomp}, the inhomogeneous effect of $k$
has less effect on the time-evolution of average value $\langle\Theta\rangle$
and $\langle\Phi\rangle$ and their fluctuations $\Delta\Theta$ and $
\Delta\Phi$; while the electron spin noise can change the magnetization
dynamics significantly. The magnetization switch time for spin polarization
vector $\mathbf{S}=(0,0,0.5)$ is about twice of that for the fully-polarized
electrons. The magnetization noise is also obviously enhanced, although its
main contribution is still from the scattering process shown in Fig.~\ref
{FeynDiagram}.

\begin{figure}[tbp]
\flushleft
\includegraphics[scale=0.32]{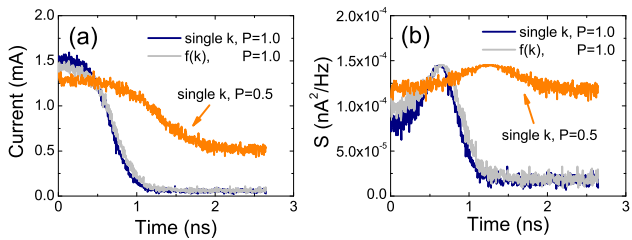}
\caption{(color online). Output electric current and current noise. (a) Current
during the switching of magnetization in Fig.~\protect\ref{noiscomp}. (b)
Current shot noise during the switching of magnetization in Fig.~\protect\ref
{noiscomp}.}
\label{currnoise}
\end{figure}

The electric current $I$ and current shot noise $S$ in a nanomagnetic junction are given as \cite{butikker}
\begin{eqnarray}
I=\frac{e^{2}}{2\pi\hbar}V\sum_{n=1}^{N_{c}}T_{n},~S=2e\frac{e^{2}}{2\pi\hbar
}V\sum_{n=1}^{N_{c}}T_{n}(1-T_{n}),  \label{currandnoise}
\end{eqnarray}
where $T_{n}$ is the transmission probability of the electron with mixed
spin states reduced from the entangled state of Eq.~(\ref{entanwav}). $N_{c}$
is the number of transport channels, estimated as $N_{c}\simeq\mathcal{A}k_{F}^{2}$.
The channel sums can be replaced, respectively, by the ensemble average $
N_{c}\overline{\delta_{\mu}}$ and variance $N_{c} (\Delta\delta_{\mu})^2$ of
the current electrons, where $\delta_{\mu}$ is a random variable given the
value 1 or 0 for a transmitted or reflected electron. By Eq.~(\ref
{currandnoise}), the time interval $\tau$ between two successive injected
electrons is given in terms of the current as $\tau=2\pi\hbar/N_{c}eV$. Then
the number of scattering events in the simulations can be converted to the
time scale of magnetization switch. Fig.~\ref{currnoise} shows the
calculated electron current $I$ and current noise $S$ associated with the
magnetization dynamics in Fig.~\ref{noiscomp}. The electric current
decreases during the magnetization switch, because the electrons are
scattered by spin-dependent potential barriers of the nanomagnet, $
\Delta_{-}=0.1$~V and $\Delta_{+}=1.3$~V, at the starting point and final
point respectively. The calculated current shot noise $S$ rise and drop in
sequence, with the maximum amplitude corresponding to the transmission
probability $T=0.5$ for electrons. The calculated current amplitude and
switching time are comparable to the experiment results \cite{nanomag}. The
amplitude of shot noise of the order of nA$^{2}/$Hz is also consistent with
the noise measurement\cite{shotnoise}. We found that the change of current
and current noise during the magnetization switch decrease with the
reduction of current electron spin polarization.

The quantum noise generated in magnetization and electric current due to the scattering
can be veiled by the thermal noise in the system at high temperature. The current shot noise
will become important when $eV>\kappa\mathcal{T}$\cite{butikker,shotnoise}, where $\kappa$ is
the Boltzmann constant and $\mathcal{T}$ is the temperature. For the bias $V=1~$mV, this
gives $\mathcal{T}<11$~K, and the shot noise was observed experimentally at $12$~K\cite{shotnoise}.
The magnetization thermal noise can be characterized by the correlator $\langle h(t)h(t')\rangle=
(2\alpha_{0}\kappa\mathcal{T}/\gamma\mathbf{M}\mathcal{V})\delta(t-t')$ for the random effective field $h(t)$\cite{brown,foros,chud},
where $\alpha_{0}$ is the Gilbert damping coefficient, $\gamma$ is the gyromagnetic ratio, $\mathbf{M}$ is the magnetization, and $\mathcal{V}$
is the volume for the nanomagnet. Then the thermal fluctuation for the rotation angle $\vartheta(t)=\gamma\int_{t}^{t+\tau}h(t')dt'$
in the time interval $\tau$ will be $\langle\delta\vartheta^{2}\rangle_{therm}=2\alpha_{0}\kappa\mathcal{T}\tau/J\hbar$,
here we used the relation $\gamma J\hbar=\mathbf{M}\mathcal{V}$. On the other hand, the quantum noise based on Eq.~(\ref{SSch}) gives
$\langle\delta\vartheta^{2}\rangle_{quant}\sim\mathcal{O}(1/J^{2})$. Thus the characteristic temperature that the quantum noise will be comparable
to the thermal noise can be roughly estimated by the relation
\begin{eqnarray}
\alpha_{0}\kappa\mathcal{T}\sim\hbar\omega_{s},\label{relation}
\end{eqnarray}
where $\omega_{s}=1/J\tau$ is a characteristic frequency related to the switch time of the nanomagnet. For the nanomagnet in Ref.~\cite{nanomag},
$\alpha_{0}\sim 0.025$, together with our simulation results above $J\tau\sim 0.1$~ns, the characteristic temperature $\mathcal{T}$ is about $3K$.
Relation (\ref{relation}) shows that the quantum magnetization noise will be more important for smaller Gilbert damping coefficient, smaller nanomagnet,
or shorter magnetization switch time.

\emph{Conclusion} The quantum trajectory method we have developed not only reproduced the semiclassical results of STT,
but also revealed the generation of quantum noise and the noise transfer between the nanomagnet and current. More quantum
phenomena in the STT-related physics are anticipated in further studies based on this quantum picture.

\emph{Acknowledgement} We acknowledge support of this work by the U. S. Army Research Office under contract number ARO-MURI
W911NF-08-2-0032 and thank W. Yang, Y.J. Zhang, H. Suhl, and I.N. Krivorotov
for valuable discussions.


\begin{thebibliography}{99}
\bibitem{STT1} J.C. Slonczewski, J. Magn. Magn. Mater. \textbf{159},
L1(1996).

\bibitem{STT2} L. Berger, Phys. Rev. B \textbf{54}, 9353 (1996).

\bibitem{mdym} D.C. Ralph and M.D. Stiles, J. Magn. Magn. Mater.
\textbf{320}, 1190 (2008).

\bibitem{ohno08pulse} T. Devolder, J. Hayakawa, K. Ito, H. Takahashi, S.
Ikeda, P. Crozat, N. Zerounian, J.-V. Kim, C. Chappert, and H. Ohno, Phys.
Rev. Lett., \textbf{100}, 057206 (2008).

\bibitem{tomita08} H. Tomita, K. Konishi, T. Nozaki, H. Kubota, A.
Fukushima, K. Yakushiji, S. Yuasa, Y. Nakatani, T. Shinjo, M. Shiraishi, and
Y. Suzuki, Applied Physics Express \textbf{1}, 061303 (2008).

\bibitem{cui10} Y.-T. Cui, G. Finocchio, C. Wang, J.A. Katine, R.A. Buhrman,
and D.C. Ralph, Phys. Rev. Lett., \textbf{104}, 097201 (2010).

\bibitem{webb} S. Garzon, L. Ye, R.A. Webb, T.M. Crawford, M. Covington, and
S. Kaka, Phys. Rev. B \textbf{78}, 180401 (2008).

\bibitem{krivo10} X. Cheng, C.T. Boone, J. Zhu, and I.N. Krivorotov, Phys.
Rev. Lett. \textbf{105}, 047202 (2010).

\bibitem{suhl} H. Suhl, \emph{Relaxation processes in micromagnetics}
(Oxford University Press, 2007).

\bibitem{foros} J. Foros, A. Brataas, G. E. W. Bauer, and Y. Tserkovnyak,
Phys. Rev. B \textbf{79} 214407 (2009).

\bibitem{chud} A. L. Chudnovskiy, J. Swiebodzinski, and A. Kamenev, Phys.
Rev. Lett. \textbf{101}, 066601 (2008).

\bibitem{slavin} A. Slavin and V. Tiberkevich, IEEE Transactions on
Magnetics \textbf{45}, 1875 (2009).

\bibitem{Cohe} W.-M. Zhang, D.H. Feng, and R. Gilmore, Rev. Mod. Phys.
\textbf{62}, 867 (1990).

\bibitem{carmichael} H. J. Carmichael, \emph{An open systems approach to
quantum optics} (Springer, Berlin, 1993).

\bibitem{wiseman} H. M. Wiseman and G. J. Milburn, \emph{Quantum Measurement
and control} (Cambridge University Press, Cambridge, 2010).

\bibitem{supp} See supplementary material for details.

\bibitem{nanomag} I.N. Krivorotov, N.C. Emley, J.C. Sankey, S.I. Kiselev,
D.C. Ralph, and R.A. Buhrman, Science, \textbf{307}, 228 (2005).

\bibitem{butikker} Ya.M. Blanter and M. B\"uttiker, Phys. Rep. \textbf{336}, 1 (2000).

\bibitem{shotnoise} K. Sekiguchi, T. Arakawa, Y. Yamauchi, K. Chida, M.
Yamada, H. Takahashi, D. Chiba, K. Kobayashi, and T. Ono, Appl. Phys. Lett.
\textbf{96}, 252504 (2010).

\bibitem{brown} W.F. Brown, Phys. Rev. \textbf{130}, 1677 (1963).
\end{thebibliography}
\end{document}